\documentclass[letterpaper, 12 pt]{article}
\usepackage[a4paper, total={6in, 8in}]{geometry}
\usepackage{graphicx} 
\usepackage{amsmath}
\usepackage{graphicx}
\usepackage{cancel}
\usepackage{lipsum}
\usepackage{cite}
\usepackage{physics}
\usepackage{amsmath,amssymb}
\usepackage{dirtytalk}
\usepackage{physics}
\usepackage{xcolor}

\title{Sharpening the Gravitational Aharonov-Bohm effect}
\author{ Akshat Pandey\footnote{apandey.physics@gmail.com} \\ \normalsize Department of Physics, Shiv Nadar Institution of Eminence \\ \normalsize Greater Noida, Uttar Pradesh-201314, India.}

\date{}

\begin{document}

\maketitle

\begin{abstract}
We study the recent gravitational analogue of the Aharonov-Bohm effect for a classical system, namely a complex scalar field. We use this example to demonstrate that the Aharonov-Bohm effect in principle has nothing to do with quantum-mechanics. We then discuss how this classical field description can be connected to the standard one particle quantum description of the Aharonov-Bohm effect.
\end{abstract}

\section{Introduction}

In 1959, Aharonov and Bohm in their seminal paper \cite{AB} proposed a phenomenon within quantum mechanics which demonstrated that potentials and field intensities are not equivalent descriptions  \cite{vaidman}. They studied the effects of an external electromagnetic potential on the quantum wave-function of a charged particle. In particular they showed that the presence of a non-zero vector potential can lead to observable effects, even for cases where the corresponding field strength vanished. Ever since, this Aharonov-Bohm (AB) effect has become a textbook example to emphasise that potentials are more fundamental than field intensities. 

Subsequently, there have been several generalisations of the Aharonov-Bohm effect. These include relativistic generalisations \cite{khalilov, rizzi, peshkin, gamboa} and the non-stationary AB effect \cite{levitov}. Recently there have been papers exploring the electric AB effect as opposed to the magnetic AB effect initially \cite{weder} and an alternative scalar electric AB effect studied by Chiaov \textit{et al.} \cite{chiaovelectric}. Gravitational analogues for the AB effect were first studied in the 1980s \cite{ford, bezerra}. More recently, following \cite{chiaovelectric} Chiaov \textit{et al.} proposed a novel model for gravitational AB effect \cite{gab}.

The usual analysis of the AB effect involves starting with a single particle wave-function in the presence of an external potential which is responsible for the AB effect. Due to reasons related to history and practice, the AB effect is thought to be a consequence of the \say{weirdness} of quantum mechanics.  

However as Wald has emphasised \cite{waldenm}, the Aharonov-Bohm effect has nothing to do with quantum mechanics \textbf{---} the same effect can be observed for a classical field. Thought of this way, the AB effect is a consequence of the non-trivial topology of the background space the field pervades in; field strengths and potentials are equivalent in any simply-connected space.

To demonstrate this, in this paper we shall study the Gravitational AB effect in a purely classical system. As an example we will analyse a toy-model classical complex scalar field that is weakly coupled to gravity. We shall show that assuming the form of the gravitational potential used in \cite{gab}, the dynamics of the classical field are influenced in a way similar to the quantum wave-function. 

Before we move on, there is a point worth mentioning. Previously Ford \textit{et al.}  had proposed a gravitational analogue model for the quantum AB effect within classical general relativity \cite{ford}. Subsequently, this was studied in the context of cosmic strings \cite{cs}.  In contrast to these articles which study a classical gravitational analogue to the electromagnetic AB effect for a quantum wave-function, in the present paper we emphasise that the AB effect, in general, is not an intrinsically \say{quantum} effect and demonstrate it for the case of the gravitational AB effect.

The paper is organised as follows. In the next section, we begin with a scalar field coupled with gravity. After obtaining a non-relativistic limit, we study the effects of the gravitational potential on the dynamics of the field, thus sketching the AB effect for a classical system. We then discuss how this classical system would give rise to the quantum wave-function description of \cite{gab}, emphasising the ubiquity of the AB effect. We end with a summary in section 3.

\section{Gravitational AB effect}

We start with the equation of motion of a minimally coupled complex scalar field in curved spacetime

\begin{equation}
    \frac{1}{\sqrt{-g}}\partial{\mu}\left( \sqrt{-g} g^{\mu \nu} \partial_{\nu} \phi \right) - m^2 \phi = 0   
    \label{1}
\end{equation}

We are working in units with $\hbar=c=1$. We will work in within the linearised gravity regime which is analogous to coupling scalar fields to electrodynamics \cite{waldgr}. Within linearised gravity

\begin{equation}
    g_{\mu \nu} = \eta_{\mu \nu} + h_{\mu \nu}
    \label{2}
\end{equation}

Here the effects of gravity are encoded in $h_{\mu \nu}$ which can be thought as a symmetric rank-2 tensor field propagating in Minkowski spacetime and interacting with other fields in it. This is also the weak field limit of GR, $\mathcal{O}(h^2)$ terms can be neglected.

Plugging equation (\ref{2}) into (\ref{1}), we get

\begin{equation}
    \eta^{\mu \nu} \partial_{\mu} \partial_{\nu} \phi - \partial_{\mu} \left( h^{\mu \nu} \partial_{\nu} \phi \right) - m^2 \phi = 0
    \label{3}
\end{equation}

In order to obtain a Newtonian limit, we choose a particular coordinate system $(\vec{x},t)$ and make use the metric corresponding that corresponds to Newtonian gravity which is

\begin{equation}
    g_{0 0} \approx (1 - 2 \Phi) 
    \label{4}
\end{equation}

with all the other components of $h_{\mu \nu}\xrightarrow{}0$. Here $\Phi = -h^{0 0}/2$   is similar to the Newtonian gravitational potential except that it is not time independent. Equation (\ref{4}) can be thought of as the \say{electrostatic} limit of linearised gravity although this is not as precise as electrostatics is \cite{lrr}.

Equation (\ref{3}) takes the form

\begin{equation}
    \partial^{\mu} \partial_{\mu} \phi - \partial_{t}h^{0 0} \partial_{t} \phi - h^{0 0} \partial^{2}_{t} \phi - m^2 \phi = 0
    \label{5}
\end{equation}

We make the following field redefinition 

\begin{equation}
    \phi (\vec{x},t) = e^{-imt}\psi (\vec{x},t)
    \label{6}
\end{equation}

Expanding equation (\ref{5})

\begin{equation}
    \partial^2_{t} \psi - 2 i m \partial_t \psi - \nabla^2 \psi - \partial_t (h^{0 0}) (i m + \partial_t) \psi - h^{0 0} (\partial^2_t - i m \partial_t - m^2      ) \psi = 0
    \label{7}
\end{equation}

We can now study the non-relativistic limit of the field $\psi$. For point particles the non-relativistic limit is obtained by taking $|\vec{p}| \ll m$. Similarly for a scalar field, the non-relativistic limit would correspond to $|\partial_t^2 \psi| \ll m|\partial_t \psi|$ and $|\partial_t \psi| \ll m| \psi|$ \cite{paddybook}.

Therefore we end up with

\begin{equation}
    -2 i m \partial_t \psi - \nabla^2 \psi - \partial_t (h^{0 0}) i m \psi - h^{0 0} m^2 \psi = 0   
    \label{8}
\end{equation}

In pure Newtonian gravity the term $\partial_t h^{00}$ vanishes but the gravitational AB effect requires a time dependent potential. We shall thus assume that $\partial_t h^{00}$ although non-vanishing, is small enough to be neglected in equation (\ref{8}).

We thus end up with

\begin{equation}
    i \frac{\partial \psi}{\partial t} = -\frac{1}{2m} \nabla^2 \psi - \frac{h^{0 0} m \psi}{2} = -\frac{1}{2m} \nabla^2 \psi + m \Phi \psi
    \label{9}
\end{equation}


We see that this equation has the same form as the Schr\"{o}dinger equation in \cite{gab}. However the interpretation is quite different. Unlike the Schr\"{o}dinger equation which in this case represents a quantum particle in a gravitational potential, equation (\ref{8}) is an equation for a classical scalar field coupled to the same potential.

We can now study the gravitational AB effect. Following \cite{gab}, we impose $\Phi(\Vec{x},t)= \Phi(t)$. This can be achieved, for example, via the Equivalence Principle in a uniformly accelerated system which at least, locally is the same for all $\vec{x}$. This ensures that the gravitational field intensity $\nabla \Phi = 0$, see \cite{gab}. 

In order to make the $\Phi$ dependence on the field dynamics explicit, we perform the following separation of variables

\begin{equation}
    \psi(\vec{x},t) = \chi (\vec{x}) \tau (t)
    \label{10}
\end{equation}

This lets us write the equation of motion of $\tau$ in the following form

\begin{equation}
     \frac{i}{\tau}\frac{d \tau}{d t} = -\frac{1}{\chi}\frac{\nabla^2 \chi}{2m} + m \Phi 
    \label{10}
\end{equation}

The $\chi$ term here is related to the gradient energy density of the field. Since it is time-independent, it is a constant in the above equation and is represented by $\mathcal{E}_{\nabla}$.

Therefore we end up with

\begin{equation}
    i \frac{d \ln{\tau}}{dt} = \mathcal{E}_{\nabla} + m \Phi
    \label {rev1}
\end{equation}

Integrating this equation with respect to $t$ we gives

\begin{equation}
    \tau(t) = \exp(- i \mathcal{E}_{\nabla} t ) \exp(-i m \int^{t}_{0} \Phi(t') dt')
    \label{rev2}
\end{equation}

The full solution becomes

\begin{equation}
    \psi(\vec{x}, t) = \chi (\vec{x}) \exp(- i \mathcal{E}_{\nabla} t ) \exp(-i m \int^{t}_{0} \Phi(t') dt')
    \label{rev3}
\end{equation}

The time evolution due to the $\mathcal{E}_{\nabla}$ exponential term is simply a periodic phase and is thus trivial. The rightmost term is responsible for the non-trivial time dependence. It is, in fact, the source of the gravitational AB effect; the dynamics of the classical field depends on the gravitational potential even though the gravitational field intensity $\nabla \Phi$ vanishes. For further details, see \cite{gab}.

\subsection*{AB Effect for the wave-function}
We briefly mention how the AB effect as evidenced in equation (\ref{rev3}) relates to the AB effect for a single particle wave-function, for a detailed discussion, see appendix. Starting with the classical field $\psi$ and its corresponding Lagrangian, upon defining the conjugate momentum, we can quantise the field by imposing the field commutation relations. Note, the gravitational potential still remains classical. Now within this quantum field theory we focus on the one particle states obtained by the action of the field (conjugate) on the vacuum

\begin{equation}
    \ket{\vec{x}} = \psi^{\dagger}(\vec{x})\ket{0}
    \label{14}
\end{equation}

These are in fact the position Eigenstates of one-particle quantum mechanics. A general state can be constructed by taking superpositions of these

\begin{equation}
    \ket{\Psi}= \int d^3x \Psi(\vec{x})\ket{\vec{x}}
    \label{16}
\end{equation}

The $\Psi(\vec{x})$ as introduced above is in fact the wave-function; the time independence is due to the Schr\"{o}dinger picture description of the quantum field. We can see that the wave function having been obtained from quantising the complex scalar field and focussing on a subspace of states, namely the one particle states. One can thus show (see appendix), that $\Psi$ in fact obeys the Schr\"{o}dinger equation

\begin{equation}
    i \frac{\partial \Psi}{\partial t} = \frac{-1}{2m} \nabla^2 \Psi + m\Phi\Psi
    \label{17}
\end{equation}

Thus equation (\ref{17}) was the starting point in the analysis of \cite{chiaovelectric, gab} and of several other AB effect studies.

\section{Discussions and Summary}

Equation (\ref{rev3}) governs the gravitational AB effect for a classical field; this tells us that there is nothing strictly quantum-mechanical about the gravitational AB effect. Additionally, we want to emphasise that the gravitational AB effect is not only possible for classical systems but is in fact generic, since gravity universally couples to everything. Further, it is the point particles are an exception to this. Considering the fields to be localised at individual points leads to simpler description of coupling to gravity; there are coupling terms between the classical field and the gravitational potential that do not show up in the coupling of a point particle to the gravitational field. However, one must note that for several reasons, the notion of point-particles in General Relativity is not well defined \cite{waldgr}. Often times, careful analyses of matter coupled to gravity involve replacing point-particles with classical fields. It is not difficult to see that as soon as one replaces point particles with classical fields, the gravitational AB effect, this also holds true for the traditional AB effect within electrodynamics.

Therefore, in more general scenarios \textbf{---} where the field intensities do not necessarily vanish or when one is working with more arbitrary gauge fields, the AB effects can be thought of interactions that vanish upon localising the classical fields to classical point particles. The simplest generalisation along these lines would be to relax the assumption about the gravitational potential being slowly evolving that was made to obtain equation (\ref{9}). It would be interesting to study the consequences of the $\partial_t h^{0 0}$ in equation (\ref{8}) which we neglected for our purposes as we wished to draw a correspondence to the standard Schr\"{o}dinger equation result.

Further, we must mention there are certain aspects of Aharonov Bohm effect for quantum systems that have no classical counterparts, particularly the non-local effects due to entanglement \cite{aharonovnl}. This should not be surprising as entanglement is a purely quantum phenomena.

To summarise, in this paper we worked out the gravitational Aharonov-Bohm effect for a classical complex scalar field. We showed an example of how the AB effect is not inherently a quantum mechanical result as is often thought about.

\section*{Appendix}

In this appendix, we explicitly show how the Schr\"{o}dinger equation for a single particle in a gravitational potential can be obtained from the classical Schr\"{o}dinger field. We begin with the Hamiltonian density corresponding to equation (\ref{9})

\begin{equation}
    \mathcal{L} = i \psi^* \partial_t \psi - \frac{1}{2m}|\nabla \psi|^2  - \frac{1}{2}m h \psi  
    \tag{A1}
\end{equation}

The conjugate momentum $\pi_{\psi}$ is

\begin{equation}
    \pi_{\psi} = \frac{\partial \mathcal{L}}{\partial (\partial_t \psi)} = i \psi^* \tag{A2}
\end{equation}

The Hamiltonian density $\mathcal{H}$ is 

\begin{equation}
    \mathcal{H} = \pi_{\psi} \partial_t \psi - \mathcal{L} = \frac{1}{2m} |\nabla \psi|^2 + \frac{1}{2}m h \psi
    \tag{A3}
\end{equation}

We now quantise the field by imposing the following quantum conditions

\begin{equation}
    [\psi(\vec{x}), \pi_{\psi} (\vec{y})] = i \delta^3 (\vec{x}- \vec{y})
    \tag{A4}
\end{equation}

Therefore the commutation relations between the fields is

\begin{equation}
    [\psi(\vec{x}), \psi^* (\vec{y})] = \delta^3 (\vec{x}- \vec{y})
    \tag{A5}
\end{equation}

The field $\psi$ is similar to the Dirac spinor in the sense that the field equation is first order in time. The field can be expanded as

\begin{equation}
    \psi(\vec{x}) = \int \frac{d^3p}{(2 \pi)^3} a_{\vec{p}} e^{i \vec{p} . \vec{x}}
    \tag{A6}
\end{equation}

where the corresponding commutation relations become

\begin{equation}
    [a_{\vec{p}}, a_{\vec{p}}^{\dagger}] = (2 \pi)^3 \delta^3(\vec{p}-\vec{q})
    \tag{A7}
\end{equation}

For this quantum field, the vacuum state $\ket{0}$ is defined by

\begin{equation}
    a_{\vec{p}}\ket{0} = 0
    \tag{A8}
\end{equation}

A one particle state $\ket{\vec{p}}$ is defined by

\begin{equation}
    a_{\vec{p}}^{\dagger}\ket{0} = \ket{\vec{p}}
    \tag{A9}
\end{equation}

The energy of the one particle state can be obtained by

\begin{equation}
    H \ket{\vec{p}} = \int \mathcal{H} \ket{\vec{p}} = \left( \frac{\vec{p}^2}{2m} + \frac{m h}{2} \right) \ket{\vec{p}}
    \tag{A10}
\end{equation}

A localised one particle state can be obtained by Fourier transforming $\ket{\vec{p}}$

\begin{equation}
    \ket{\vec{x}} = \int \frac{d^3p}{(2 \pi)^3} a_{\vec{p}}^{\dagger} e^{- \vec{p}. \vec{x}} \ket{\vec{p}}
    \tag{A11}
\end{equation}

This is in fact

\begin{equation}
    \psi^{\dagger}\ket{0} = \ket{\vec{x}} 
    \tag{A12}
\end{equation}

A general state can be constructed by superposing the one particle position Eigenstates

\begin{equation}
    \ket{\Psi} = \int d^3 x \Psi (\vec{x}) \ket{\vec{x}}
    \tag{A13}
\end{equation}

$\Psi$ here is the wave-function. Position and momentum operators corresponding to the one-particle state can be defined

\begin{equation}
    \vec{P} = \int \frac{d^3 p}{(2 \pi)^3} \vec{p} a_{\vec{p}}^{\dagger} a_{\vec{p}}
    \tag{A14}
\end{equation}

and

\begin{equation}
    \vec{X} = \int \frac{d^3 p}{(2 \pi)^3} \vec{x} a_{\vec{p}}^{\dagger} a_{\vec{p}}
    \tag{A15}
\end{equation}

Acting on the general state $\ket{\Psi}$ we get

\begin{equation}
    \vec{X} \ket{\Psi} = \int d^3x \vec{x} \Psi (\vec{x}) \ket{\vec{x}}
    \tag{A16}
\end{equation}

For the momentum operator

\begin{equation}
    \vec{P} \ket{\Psi} = \int \frac{d^3x d^3p}{(2 \pi)^3} \vec{p} a_{\vec{p}}^{\dagger} a_{\vec{p}} \Psi (\vec{x}) \psi^{\dagger}(\vec{x}) \ket{0} = \int \frac{d^3x d^3p}{(2 \pi)^3} \vec{p} a_{\vec{p}}^{\dagger} e^{- \vec{p}. \vec{x}} \Psi (\vec{x})  \ket{0}
    \tag{A17}
\end{equation}

Combining $\vec{p}$ and $e^{- i \vec{p}. \vec{x}}$ into a derivative term we get

\begin{equation}
    \vec{P} \ket{\Psi}= \int \frac{d^3x d^3p}{(2 \pi)^3} e^{- i \vec{p}. \vec{x}} \left(-i \nabla \Psi \right) a_{\vec{p}}^{\dagger} \ket{0} = \int d^3x \left( -i \nabla \Psi    \right) \ket{\vec{x}}
    \tag{A18}
\end{equation}

We see that $\vec{X}$ and $\vec{P}$ satisfy the non-relativistic particle commutaion relations 

\begin{equation}
    [X^i, P^j] = i \delta^{ij} \ket{\Psi}
    \tag{A19}
\end{equation}

Similarly the one particle Hamiltonian operator is defined as

\begin{equation}
    H = \int d^3x \frac{1}{2m} |\nabla \psi|^2 + \frac{1}{2}m h \psi = \frac{1}{2} m h \int \frac{d^3p}{(2 \pi)^3} a_{\vec{p}} e^{i \vec{p} . \vec{x}} + \int \frac{d^3p}{(2 \pi)^3} \frac{\vec{p}^2}{2m} a_{\vec{p}}^{\dagger} a_{\vec{p}} 
    \tag{A20}
\end{equation}

Acting on $\ket{\Psi}$ we end up with

\begin{equation}
    i \frac{\partial \Psi}{\partial t} = \frac{-1}{2m} \nabla^2 \Psi + m\Phi\Psi
    \tag{A21}
\end{equation}

We thus end up with the one-particle Schr\"{o}dinger equation placed in a gravitational potential.


\end{document}